\documentclass[a4paper]{article}

\usepackage[dvips]{graphicx}

\title{Surviving opinions in Sznajd models on complex networks}

\author{F. A. Rodrigues and L. da F. Costa}

\begin{document}
\maketitle

\section{Abstract}

The Sznajd model has been largely applied to simulate many
sociophysical phenomena. In this paper we applied the Sznajd model
with more than two opinions on three different network topologies and
observed the evolution of surviving opinions after many interactions
among the nodes. As result, we obtained a scaling law which depends of
the network size and the number of possible opinions. We also observed
that this scaling law is not the same for all network topologies,
being quite similar between scale-free networks and Sznajd networks
but different for random networks.

\section{Introduction}

Von Neumann and Ulam introduced the concept of cellular automata
in the early 1950's. Ever since, this concept has attracted
continuing interest and has been subject to deep mathematical and
physical analysis. A good deal of the popularity of cellular
automata arises from their simplicity and potential to model many
complex systems. The applications of this theory range from
modelling biological pattern formation \cite{Alt:1997} to
sociophysical phenomena \cite{Stauffer:2003}.

In sociophysics, attention has been focused on modeling social
phenomena, such as elections and propagation of information
\cite{Stauffer:2003}. With this respect, particularly successful
models have been developed by Sznajd-Weron \cite{Sznajd:2000},
Deffuant et. al \cite{Deffuant:2000} and Krause and Hegselmann
\cite{Hegselmann:2002}, which differ as for their definitions but
tend to produce similar results.

Among those three models, the one developed by Sznajd is the most
appropriate for simulation in networks and lattices, because it
considers just the interactions between the nearest neighbors. The
Sznajd model has been developed based on Ising model and been
successfully applied to model sociological and economics systems
\cite{Sznajd:2005}. In this paper we simulate the Sznajd model on
networks with different topologies.  The results turned out to be
dependent on the network topology only for small values of the
ratio between the number of possible opinions and the network
size.

\section{Sznajd model on complex networks}

Complex networks are formed by a set of nodes ($i = 1, 2... N$),
which are linked one another through edges. In biological
networks, for example, proteins or genes can be linked according
to their interactions \cite{Barabasi:2004}. In case of social
networks, one can represent connections as defined by human
relations such as friendship \cite{Rapoport:1961}, relations
between jazz musicians \cite{Gleiser:2003}, collaborations in
scientific researchers networks\cite{Barabasi:2002} and
intermarriage between families \cite{Padgett:1993}.

To explain real network topologies, some models of complex
networks have been developed, including random graphs
\cite{Bollobas:1985,Erdos-Renyi:1959}, small-world networks
\cite{Watts:1998}, scale-free networks \cite{Barabasi:2000} and,
more recently, Sznajd complex networks \cite{Costa:2005}.

As complex networks offer a more structured and realistic topology
than regular lattices, the simulation of sociophysics models in
these networks can produce more accurate results
\cite{Stauffer:2004b}. Stauffer at. al. simulated the Deffuant et.
al. model in scale-free networks \cite{Stauffer:2004a}, obtaining
the result that different surviving opinions in scale-free
networks depend on the network size and the number of possible
opinions. Moreover, Bernardes et. al \cite{Bernardes:2002},
developed a model using the Sznajd model on a scale-free network
which reproduced Brazilian election results.  It was shown that
the distribution of number of votes per candidate follows a
hyperbolic law, in agreement with real elections results. Thus,
simulation of sociophysical models on complex networks structures
can reproduce real phenomena and suggest insights on the rules
that guide de distribution of opinion dynamics.

\subsection{Models of networks}

The random graph is the simplest complex network model. The number
of nodes on random graphs is constant and new edges randomly
connect them with probability $p$. Thereby, the distribution of
connections in the network will follow the Poisson distribution.
Thus, the node degree, given by the number of connections of each
node, of the majority of network nodes will be next to the
average.

In the case of scale-free networks, recently added nodes have
greater probability to connect with those which have more
connections. This process, called \emph{preferential attachment},
generates a heterogeneous network, where most nodes will have a
few number of connections, while a few nodes will a have high
number of connections. The connectivity distribution, $P(k)$, for
this network follows a power law distribution, $P(k) \sim
k^{-\gamma}$, where $k$ is the number of links.

\begin{figure}[h]
\begin{center}
  \centerline{\includegraphics[width=10cm]{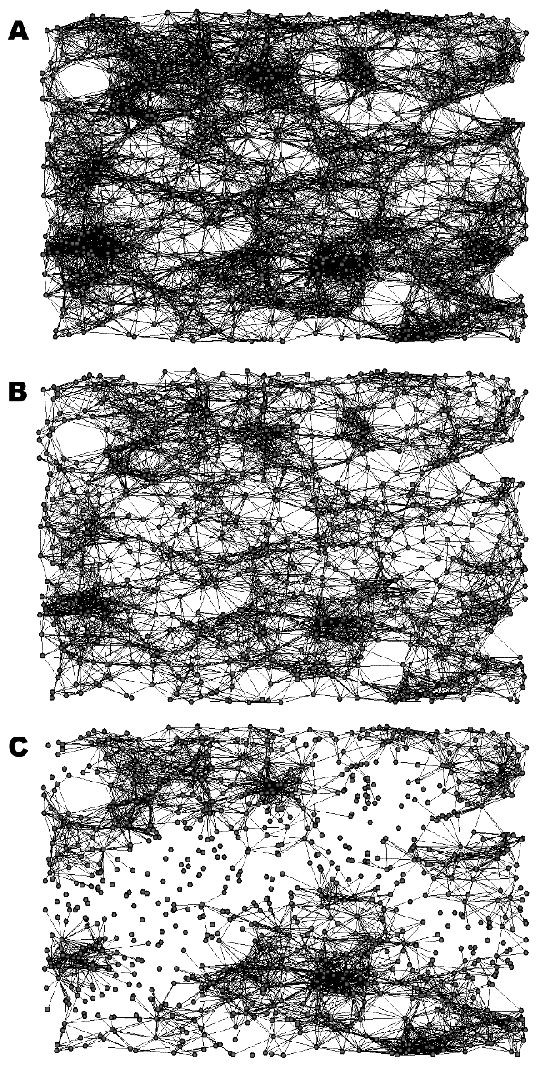}}
  \caption{(a) The construction of Sznajd network starts with a geographical distribution
  of nodes at random and linking them that are distant less than a defined euclidian distance $d_{max}$.
  (b) Then, each of the network links is activated with probability $p$. (c) Finally,
the feedbakc Sznajd and contrary feedback are applied, thus
resulting the Sznajd network.}
 \label{sznajd}
\end{center}
\end{figure}

The Sznajd network is constructed by considering the geographical
distribution of nodes and the evolution of Sznajd dynamics with
feedback (contras) \cite{Costa:2005}. In this way, the network
starts with a constant number of nodes distributed in a box of
side $L$ at random. Then, nodes distant less than $d_{max}$ are
connected in order to obtain the network $U$ as shown in Figure
\ref{sznajd}(a). Then, the edges of this network are activated
with uniform probability $p$, thus obtaining a network $K$, as
shown in Figure \ref{sznajd}(b).  At each step, an edge $(i,j)$ is
sampled at random from the initial network $U$. If the edge is
present in the network $K$, all neighbors of $i$ and $j$ are
identified in $U$ and connected in $K$. Otherwise, all neighbors
of $i$ and $j$ are disconnected in $K$. Another edge of $U$ is
sampled with probability $q$ and the respective edge in $K$
receives the contrary value to the current dominant opinion in the
network. After stabilization, the geographical Sznajd network is
obtained and it is observed the presence of network communities,
formed by sets of densely connected nodes.

\begin{figure}[h]
\begin{center}
  \centerline{\includegraphics[width=10cm]{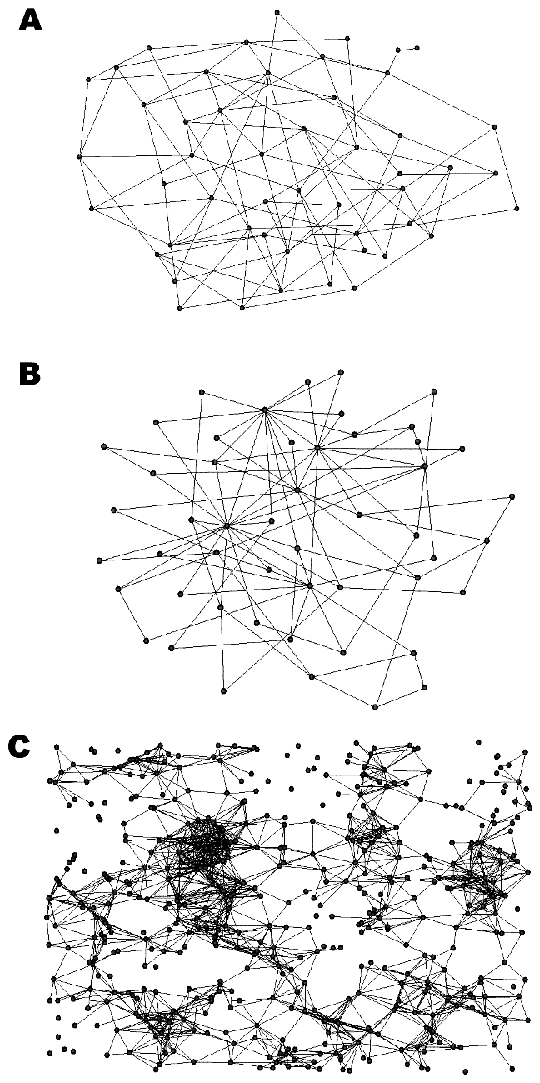}}

  \caption{The simulation of Sznajd model is carried out for three different
  topologies: (a) random graphs,(b) scale-free networks and (c) Sznajd
  networks. }
 \label{models}

\end{center}
\end{figure}

Examples of random network, scale-free networks and Sznajd
networks are presented in Figure \ref{models}. In our simulation,
we used these three topologies to analyze the distribution of
surviving opinions while varying the size of the network and the
number of possible opinions.

\subsection{Simulations}

The most widespread version of the Sznajd model uses a square
lattice with two opinions, $Q = \pm 1$ \cite{Stauffer:2000}, where
each individual $i$ ($i = 1,2,...,N$) has equal number of
neighbors. The simulation of this model starts by distributing
opinions at random on the lattice. Then, at each step, two
neighbor pairs are selected and if they have the same opinion all
their neighbors are made to agree with them. If the initial state
of the system has more than half opinions as $1$, the final
consensus is reached with all individuals reaching this opinion.
Thus, a phase transition is observed \cite{Stauffer:2000}.
Nevertheless, such a model is very simple and cannot reproduce
some real results. To overcome the limitations, improvements were
added to the Sznajd model in order to consider more than two
opinions ($q = 1,2... Q$) as well as more sophisticated real
topologies, including complex networks
\cite{Bernardes:2002,Stauffer:2004}.

The simulation of the Sznajd model on complex networks is similar
to performed on lattices: a pair of neighboring nodes are chosen
at random and checked if they have the same opinion. If they do,
all their nearest neighbors assume that same opinion.

\section{Results}

For the three models of networks (i.e. random graph, scale-free
network and Sznajd network), $Q$ different opinions were randomly
distributed initially among the nodes. The interaction between
nodes proceeds by choosing uniformly a node and one of its
neighbors at random. Nobody can convince anyone
\cite{Deffuant:2000,Hegselmann:2002} if the two opinions differ by
more than $|\epsilon|$; in our simulation we adopted
$\epsilon~=~1$. So, if two neighbor nodes have opinion $q~=~2$,
they can convince just their neighbor with opinions $q~=~1$ or
$q~=~3$.  This process was executed for several numbers of
opinions, varying from $2$ to $Q$, and the number of surviving
opinions was recorded at each step for different network sizes.

\begin{figure}[h]
\begin{center}
  \centerline{\includegraphics[width=8cm]{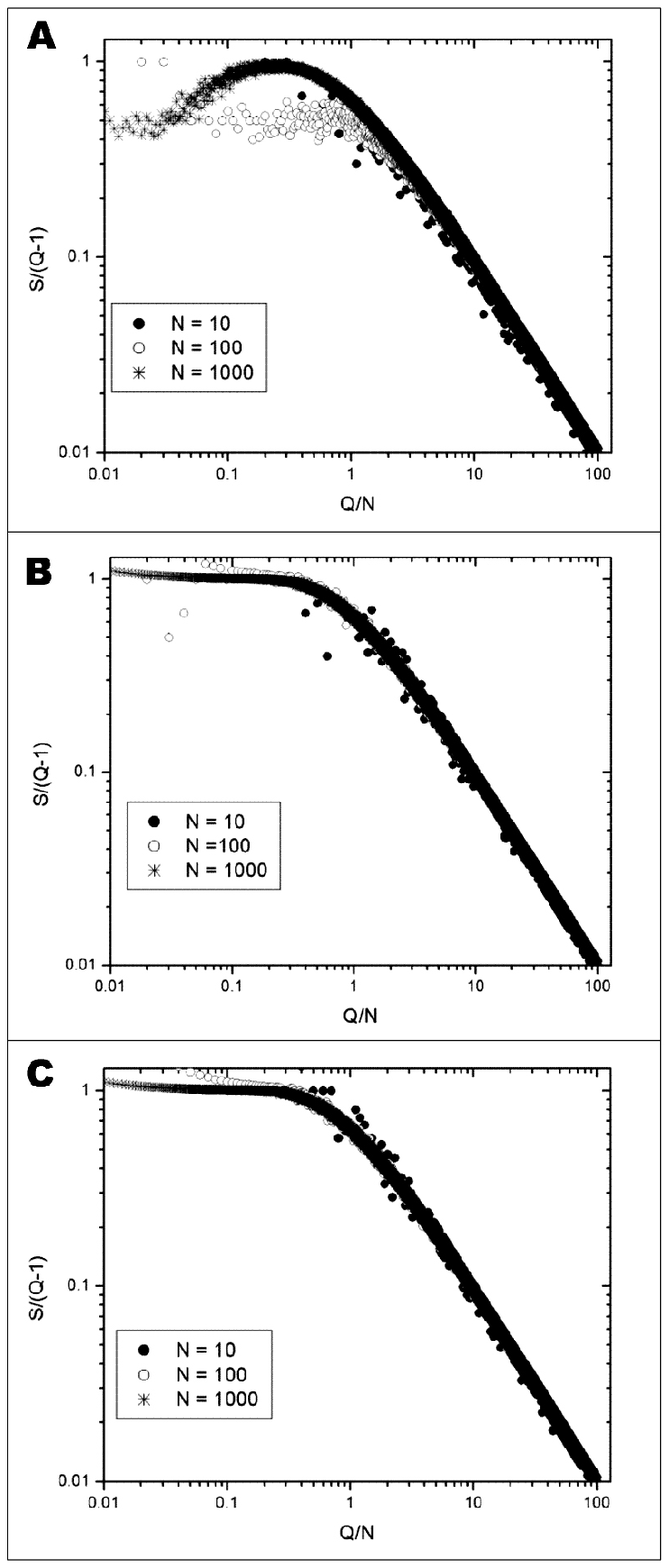}}
  \caption{Scaling of the number $S$ of surviving opinions for $N = 10, 100, 1000$ as a
  function of the number of possible opinions. We see at the right part of the graph,
  when $N\gg Q$, all person keep his/her own opinion. On the other
  hand, at the left hand side of the graph, $Q \gg N$, each opinion is shared by many people.}
 \label{results}
\end{center}
\end{figure}

In Figure \ref{results} it is shown the scaling behavior of the
number of surviving opinions as a function of the $Q$ possible
opinions. In fact, if the number of people, $N$, is much larger
than the number of possible opinions, $Q$, the number of surviving
opinions, $S$, will tend to agree with $Q$, and no opinion will
disappear during the process. However, when the number of opinions
$Q$ is much larger than the number of people $N$, the number of
surviving opinions $S$ will become $N$, i.e. each person keeps its
own opinion. This relation can be mathematically expressed as
\begin{equation}  \label{eq}
\frac{S}{Q} = f(\frac{Q}{N}),
\end{equation}
where $f$ is constant for $Q\ll N$ and $f = N/Q$ for $Q\gg N$,
valid for large $Q,S$ and $N$ \cite{Stauffer:2005}.

\section{Discussion}

The obtained results are similar to those considering the Deffuant
model on a scale free network \cite{Stauffer:2004a}.  Therefore,
the distribution of surviving opinions for Sznajd simulation
yields two limits: (i) when there are many people and few
opinions, all opinions have some followers and (ii) when there are
few people and many opinions, each person will keep her/his own
opinion. The interval between these two extremes follows a scaling
law.

As shown in Figure \ref{results}, the scaling number of surviving
opinions depends on the network topology. For random graph the
scaling of surviving opinion is not well defined for $N \gg Q$.
However, for scale-free network and Sznajd networks, the number of
surviving opinions is well determined for larger networks.  As
social networks are not guided by random distribution of
connections \cite{Newman:2003}, the Sznajd model is more likely to
reproduce dynamical behavior of opinions in real social networks.
Note that the scale-free and Sznajd network models implied similar
opinion dynamics, while the random model produced results which
originates from Equation~\ref{eq} for small values of $Q/N$.

\section{Conclusion}

In this paper we simulated the Sznajd model in three different
network topologies, namely random, scale-free and Sznajd network
models. Starting from a network with $N$ agents and varying the
number of possible opinions from $2...Q$, we simulate the Sznajd
interaction between the agents and calculated the number of
surviving opinions after a large number of interactions. As
result, we obtained that the number of surviving opinions
undergoes two states: (i) when the number of agents is much larger
than the number of opinions, the number of surviving opinions
tends to be the same as for possible opinions, where all opinions
have some followers; and (ii) when the number of possible opinions
is much larger than the number of agents, the number of surviving
opinions will tend to that number of agents, and each person will
keep her/his own opinion.

The behavior described by Equation~\ref{eq} is best fitted by
scale-free networks and Sznajd networks, where the distribution of
connections between nodes is not uniform. For random graphs, this
behavior is valid just when $Q > N$. As real social networks are
not described by random graphs, the Sznajd model is appropriate to
model dynamical behavior of opinions on this kind of networks.

\section{Acknowlegments}

Luciano da F. Costa is grateful to FAPESP (proc. 99/12765-2), CNPq
(proc. 308231/03-1) and the Human Frontier Science Program
(RGP39/2002) for financial support. Francisco A. Rodrigues
acknowledges FAPESP sponsorhip (proc. 04/00492-1).

\end{document}